\documentclass[10pt,journal,compsoc]{IEEEtran}
\ifCLASSOPTIONcompsoc
  \usepackage[nocompress]{cite}
\else
  \usepackage{cite}
\fi
\ifCLASSINFOpdf
\else
\fi
\usepackage[numbers]{natbib}
\usepackage{booktabs} 
\usepackage{amsmath, mathtools, cleveref}
\usepackage{ amssymb }
\usepackage{multirow}
\usepackage{wrapfig}
\usepackage{subfigure}
\usepackage{mathtools}
\usepackage{color}
\usepackage{rotating}
\usepackage{array,url}
\usepackage{balance}
\usepackage{tabularx}
\usepackage{epstopdf}
\usepackage{subfigure}

\begin{document}
%
\title{Speech-Driven Expressive Talking Lips with Conditional Sequential Generative Adversarial Networks}
%
%
%
%

\author{Najmeh~Sadoughi
        and~Carlos~Busso
\IEEEcompsocitemizethanks{\IEEEcompsocthanksitem N. Sadoughi and C. Busso are with the Erik Jonsson School of Engineering \& Computer Science, The University of Texas at Dallas, Richardson TX 75080.\protect\\
E-mail: najme.sadoughi@gmail.com, busso@utdallas.edu}
\thanks{Manuscript submitted to IEEE Transactions on Affective Computing}
}

\IEEEtitleabstractindextext{%
\begin{abstract}
Articulation, emotion, and personality play strong roles in the orofacial movements. To improve the naturalness and expressiveness of \emph{virtual agents} (VAs), it is important that we carefully model the complex interplay between these factors. This paper proposes a conditional generative adversarial network, called \emph{conditional sequential GAN} (CSG), which learns the relationship between emotion and lexical content in a principled manner. This model uses a set of articulatory and emotional features directly extracted from the speech signal as conditioning inputs, generating realistic movements. A key feature of the approach is that it is a speech-driven framework that does not require transcripts. Our experiments show the superiority of this model over three state-of-the-art baselines in terms of objective and subjective evaluations. When the target emotion is known, we propose to create emotionally dependent models by either adapting the base model with the target emotional data (CSG-Emo-Adapted), or adding emotional conditions as the input of the model (CSG-Emo-Aware). Objective evaluations of these models show improvements for the CSG-Emo-Adapted compared with the CSG model, as the trajectory sequences are closer to the original sequences. Subjective evaluations show significantly better results for this model compared with the CSG model when the target emotion is happiness.
\end{abstract}

\begin{IEEEkeywords}
Speech-driven model, lip movements, expressive and naturalistic lip movements, generative adversarial network.
\end{IEEEkeywords}}

\maketitle

\IEEEdisplaynontitleabstractindextext

%

\IEEEraisesectionheading{\section{Introduction}\label{sec:introduction}}
\label{sec:intro}
\IEEEPARstart{T}
{he} orofacial area  plays a role in conveying lexical, emotional and idiosyncratic information. These factors  are integrated in a nontrivial manner, facilitating face to face communications. It is important to generate proper facial movements for \emph{virtual agents} (VAs) to communicate a message more effectively and more naturally. Although emotion is expressed throughout the whole face, there are emotional states such as happiness for which the orofacial area plays a big role (e.g., smile). For these emotions, in particular, careful modeling of the relationship between emotion and articulation is required to have more natural and expressive VAs. 

Several factors contribute to the variation in the orofacial area. The orofacial muscles are activated by the articulatory movements imposed through the vocal region. The relation between lip motion and phonetic content is colored by the emotional cues expressed in the message. This coupling between lexical and emotional contents is also affected by idiosyncratic characteristics across people. The integration between these factors in the orofacial area is complex \cite{Mariooryad_2013, Busso_2006}. Most of previous studies on lip movement synthesis have relied on the recordings from one subject in order to avoid speaker variations \cite{Taylor_2016, Karras_2017, Anderson_2013}. Since multimodal emotional corpora usually include multiple speakers with limited data per subject \cite{Li_2016_2}, it is important that the models can effectively capture speaker variability. If these variations are not carefully considered, the model may predict trajectories that average these variations, creating over-smoothed movements. Furthermore, most of previous models for lip movements rely on transcriptions (e.g., phonemes or tri-phonemes) \cite{Taylor_2017,Xu_2013,Deng_2005_2}, or transcriptions plus the target emotional categories \cite{Cao_2005, Parker_2017, Anderson_2013}. The need for transcriptions limits the domain of applications. We envision a data-driven lip generation framework that does not require transcription, and can effectively capture the temporal relations between speech, lip movement and emotion.

Speech conveys verbal and nonverbal cues, having a direct influence in the visual appearance of the orofacial area. For example, speech is one of the primary channels to convey emotions \cite{Lee_2014_2}. Therefore, relying on speech for modeling the nonverbal behaviors in the orofacial area can help the model to capture the fine expressive movements shown during natural interactions. Our envisioned framework relies on speech features to generate lip motion. 

This paper proposes to use a \emph{conditional generative adversarial network} (cGAN), composed of \emph{long short-term memory} (LSTM) for generating realistic and expressive lip movements. The approach is called \emph{conditional sequential generative adversarial networks} (CSG). The model learns the distribution of the orofacial movements conditioned on speech features. The training of the models consists of an adversarial objective that combines a generator and a discriminator, such that it generates more convincing lip movements. A key  feature of the adversarial training is to teach the model to capture the temporal relationship between acoustic features and lip motion. This objective is achieved by creating fake sequences with mismatched speech and lip motion trajectories that the discriminator has to recognize. The resulting lip motion sequences capture the temporal coupling between speech and lip movements, creating realistic sequences, which convey the underlying lexical content. We compare the CSG model with three baselines proposed in previous studies \cite{Sadoughi_2017,Taylor_2016, Fan_2016}, which use conventional non-adversarial methods. The experimental evaluations with objective and subjective metrics demonstrate that the proposed CSG model achieves better performance than these methods.

Another appealing property of the CSG framework is that it can be easily extended to consider the target emotional category of the test sample during the generation. We build two expression-aware models: (1) by adapting the CSG model to different emotions, called CSG-Emo-Adapted model, and (2) by conditioning the CSG model on categorical emotion of the speaker, called CSG-Emo-Aware model. Objective evaluations show that the expression-aware models generate more expressive orofacial movements. Subjective evaluations show that the CSG-Emo-Adapted model generates better expressions compared to the CSG model, when the target emotion is happiness. The results validate our proposed method, which can generate more convincing and expressive orofacial movements.

The rest of the paper is organized as follows. Section \ref{sec:relatedwork} reviews related work. Section \ref{sec:resources} describes the resources used in this study including the corpus, the extracted audiovisual features, and the rendering toolkit. Section \ref{sec:methodology} describes the \emph{conditional sequential GAN} (CSG) method, and its two expression-aware extensions: the CSG-Emo-Adapted and CSG-Emo-Aware models. Section \ref{sec:ExperimentalEvaluation} describes the experimental evaluation. Section \ref{sec:results} presents the results of the models, comparing the CSG methods with the baselines. Section \ref{sec:conclusion} summarizes the contributions of this work highlighting the advantages and disadvantages of the approach, and possible future directions.

\section{Related Work}
\label{sec:relatedwork}

This section summarizes  previous studies to generate lip motion. We group these studies into three categories: unit selection, \emph{hidden Markov model} (HMM)-based approaches, and \emph{deep neural network} (DNN)-based approaches. 

\subsection{Unit Selection}
\label{ssec:unit}

Unit selection methods consists of selecting canonical shapes which are blended. The weights depend on the underlying phonetical unit. \citet{Xu_2013} considered several canonical shapes for lips, where the weights for phonemes and bigrams were carefully defined by artists. \citet{Deng_2005_2} modeled coarticulation between phonemes, relying on real recordings of human data. They found the weights for the linear combinations of the canonical shapes for diphones and triphones, minimizing the error between the predicted and the original movements. \citet{Cao_2005} developed a framework to generate expressive facial movements. Their framework used tuples containing phoneme, emotion, prosody, and lip trajectories. During testing, the input was parsed with the phonetic content, and the target or predicted emotion. The database was searched with the sequence of tuples derived from the input, while imposing correct co-articulation and smooth constraints. The selected segments were aligned with the input using time-warping. Finally, the motion segments were blended and smoothed to create facial movements. 

Unit selection methods require emotion-dependent speech units to account for expressive lip motions, where the weights have to be redefined for each of the target emotion. 

\subsection{HMM-based Modeling}
\label{ssec:gen}
	
	HMM-based models learn to synthesize lips movements from text or speech by implicitly modeling the underlying co-articulations. \citet{Choi_2001} proposed to use HMM inversion for audio to visual conversion. They used a three-state HMM to model each phoneme. During testing, they relied on the Baum-Welch algorithm to find the maximum likelihood estimates of the visual features. \citet{Xie_2007_2} proposed to use \emph{coupled HMMs} (CHMMs) to model the dependencies as well as the differences between the audio and visual modalities (e.g., their asynchrony and different number of phonetic units). \citet{Anderson_2013} designed a system to create emotional facial movements using \emph{cluster adaptive training} (CAT), which was built upon HMMs for text-to-speech systems. Their HMM modeled quinphones created with five states, where a decision tree was used to handle the sparseness of the quinphones in the data. The decision tree is also used to find the mean and variances of the Gaussian distributions for the quinphone. The proposed CAT framework captured emotion dependent quinphones by finding emotion-dependent linear combinations between clusters.

\subsection{DNN-based Modeling}
\label{ssec:dis}

DNN-based models directly learn how to predict the movements from speech features. \citet{Taylor_2016} proposed a fully connected feedforward neural network for audio to visual conversion. Their network gets the speech features over a specified contextual window, predicting current and future orofacial movements. The approach used sliding windows with step size of one frame where the average of the predictions for each window is considered as the target value for the center of the window. Their model outperformed the HMM inversion approach proposed by \citet{Choi_2001}. \citet{Fan_2016} explored the use of deep learning structures built with \emph{bidirectional long short-term memory} (BLSTM) to synthesize head and face movements driven by transcriptions, speech, and transcriptions plus speech. The inputs of the system correspond to triphoneme labels from transcriptions, and/or \emph{mel frequency cepstral coefficients} (MFCCs) and their first and second order derivatives from speech. The study compared the results achieved with their model with a HMM-based approach, showing improvements in terms of objective and subjective metrics.  \citet{Li_2016_2} proposed strategies using BLSTM models to create emotional facial movement by having access to a small emotional dataset. They proposed several approaches to leverage recordings from a neutral corpus and a small emotional corpus, aiming to improve the emotional regression result. Their best result was achieved with a cascade framework, where the predictions obtained with the neutral corpus are concatenated with audio features and used as feature of a second system. The second system is trained with the emotional corpus.

\citet{Karras_2017} proposed a framework with \emph{convolutional neural networks} (CNNs) to predict facial movements from raw speech signal. Their framework disentangled the facial configurations explained by audio features and emotional states. This goal is achieved by considering a dedicated emotional state learned for each training sentence. Their framework predicted the facial pose for one frame at a time, utilizing a contextual window of 260ms with previous and future frames. They trained separate models per speaker with three to five minutes of synchronized audiovisual data. They compared their method with the results achieved by the faceFX software using subjective evaluations showing higher preferences for their models. \citet{Parker_2017} proposed an approach for generating emotional audiovisual content from transcriptions and target emotion. They proposed to share the layers of the network across all the emotions, with the exception of the last layer, which was adapted for each emotion using regularized least squares. They compared their results with an HMM system, showing improvements when using their method.

\subsection{Contributions}
\label{ssec:contributions}

This papers proposes to use a conditional GAN structure composed of BLSTM units called CSG to learn the distribution of orofacial movements conditioned on speech features. The proposed CSG framework relies on adversarial training by jointly training a generator and a discriminator. During the adversarial training, a discriminator learns to recognize two sets of fake samples: the samples generated by the generator, and samples from uncoupled recordings from the original database where the audio does not match the lip motion sequence. As the generator learns to create realistic sequences to fool the discriminator, our method generates realistic samples which are timely coupled with the audio. To the best of our knowledge, this is the first time that adversarial training is used to synthesize lip motion. By using conditional GAN, we effectively model the relationship between emotion, speech and orofacial movements. This framework departs from deep learning approaches used by previous studies, providing a systematic strategy to generate emotional lip sequences. 

\section{Resources}
\label{sec:resources}

\subsection{The IEMOCAP Corpus}
\label{ssec:iemocap}

This study uses the IEMOCAP corpus \cite{Busso_2008_5}. This database comprises video, audio, and motion capture recordings from 10 actors in improvised and script-based scenarios. The scenarios were designed such that they elicited different emotions from the actors. We use the data from all the subjects, where 60\% of the data is used for training, 20\% for validation and 20\% for testing. The database is annotated with categorical emotions by three annotators at the speaking turn level. They annotated the emotional content using ten classes: neutral state, anger, happiness, sadness, surprise, fear, frustration, excitement, disgust, and other. Similar to previous studies relying on the IEMOCAP corpus, we merge the turns labeled with excitement and happiness \cite{Metallinou_2010, Mariooryad_2016}. We calculate the consensus labels for each turn by estimating the majority vote across the annotations. This approach creates hard emotional classes for each sentence. The frequencies of emotional categories for the consensus labels are 605 (neutral state), 621 (anger), 882 (happiness), 653 (sadness), 1 (disgust), 20 (fear), 998 (frustration), 31 (surprise), and 2 (other). The evaluators do not reach agreement in 1,228 segments. Due to the sparsity of the classes disgust, fear, and surprise, we merge all these segments with the speaking turns without consensus, assigning them to the class other. Consensus labels such as majority vote discard information provided by individual evaluations (see study by  \citet{Lotfian_2017}). Therefore, we also rely on soft assignments by considering the individual annotations (three annotations per turn). The frequencies of the emotional classes assigned to the turns when we consider individual annotations are: 2,538 (neutral state), 2,108 (anger), 3,795 (happiness), 2,047 (sadness), 55 (disgust), 138 (fear), 3,961 (frustration), 200 (surprise) and 281 (other). For consistency, we restrict the analysis to the six classes neutral state, anger, happiness, sadness, frustration, and other. The soft labels are created by estimating the distribution of the labels assigned to the speaking turn. For example, if there are two annotations for anger and one for frustration, we consider a 6D vector with 0.66 for anger, 0.33 for frustration and 0.0 for the remaining categories.

\subsection{Audiovisual Features}
\label{ssec:features}

We extract two sets of features from the audio. The first set of features are 25 MFCCs extracted with Praat \cite{Boersma_1996} over 25ms windows every 8.33 ms. We choose 25 MFCCs, because  \citet{Taylor_2016} evaluated with different number of MFCCs for predicting lip movements, finding that 25 MFCCs gives the best result. By moving the analysis window in increments of 8.33ms, we create 120 feature vectors per second, matching the sampling rate of the motion capture recordings. We also extract the fundamental frequency and intensity with Praat using the same window and step size. Moreover, we extract 17 additional \emph{low level descriptors} (LLDs) from the \emph{extended Geneva minimalistic acoustic parameter set} (eGeMAPS) \cite{Eyben_2016}, which is a feature set carefully selected for paralinguistic tasks. The eGeMAPS features are extracted with OpenSmile \cite{Eyben_2010_2}.

\begin{figure}[t]
	\centering
	\subfigure[Markers - IEMOCAP corpus]{
		\includegraphics[width=\columnwidth/3]{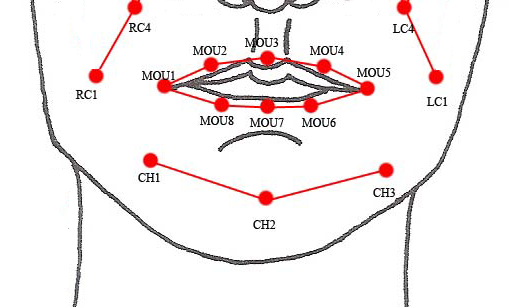}
		\label{fig:markers}
	}
	\subfigure[Markers used by Xface]{
		\includegraphics[trim = 8cm 10.2cm 12cm 2cm, clip, width=4cm]{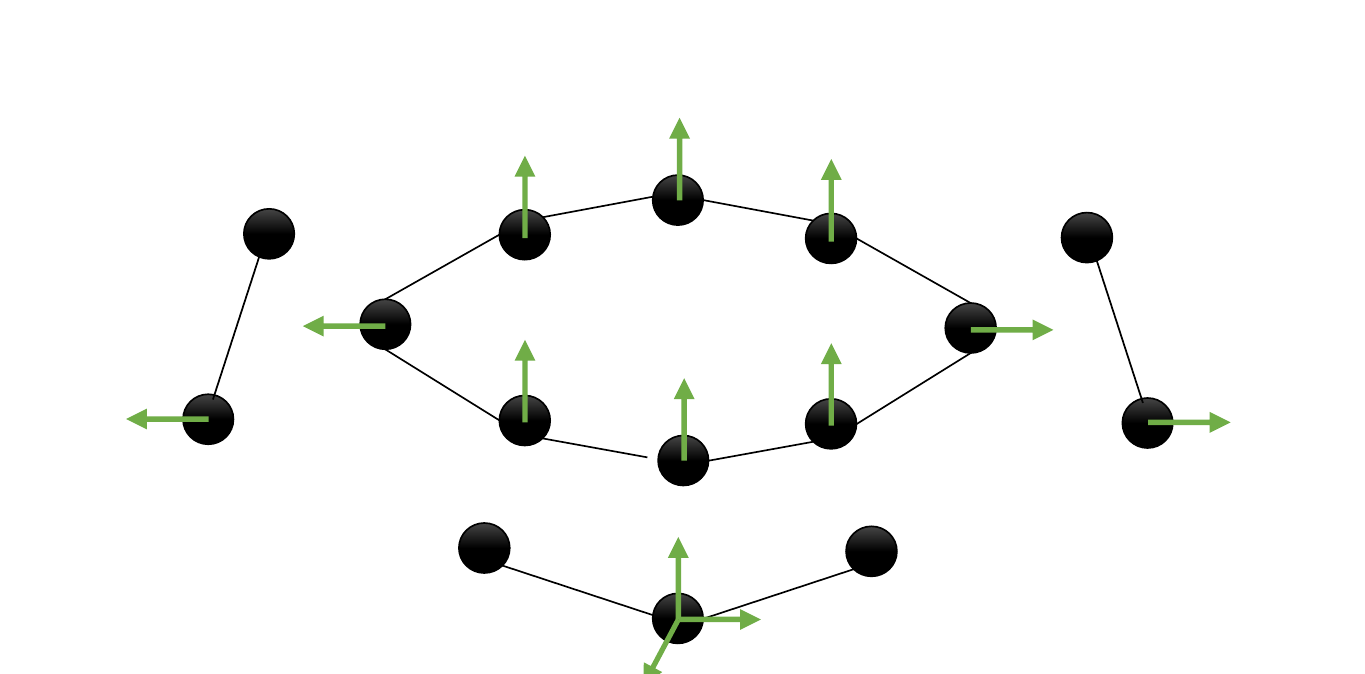}
		\label{fig:mapping-xface}		
	}
	\caption{The location of the 15 markers from the IEMOCAP corpus considered in this study. These markers are used to render the animations using Xface.}
	\label{fig:IEMOCAP}
\end{figure}

From the motion capture recordings, we use the $(X,Y,Z)$ locations of 15 markers around the mouth area (Fig. \ref{fig:IEMOCAP}). The sampling rate is 120 fps.  \citet{Busso_2008_5} describes the steps to derive the motion capture data.  

\subsection{Xface}
\label{ssec:xface}    
We rely on Xface \cite{Balci_2004} for rendering the VA. Xface uses \emph{facial action parameters} (FAPs) to animate the face. FAPs control  \emph{facial points} (FPs) on the face which are based on the MPEG-4 standard. Most of the facial markers in the IEMOCAP corpus follow the locations of FPs in the MPEG-4 standard (Fig. \ref{fig:mapping-xface}), so it is possible to linearly map the markers and FAPs. This mapping is achieved by mapping a neutral pose of the recording of each subject as a the reference pose. Then, we map the range of movements for each marker to the range of movements allowed by the FAPs in Xface. More details about this mapping is provided by \citet{Mariooryad_2012_2}. This study uses a female character for all the subjective evaluations. 

While there are other more sophisticate rendering toolkits, Xface allows us to easily animate the motion capture data in our corpus. As a result, we can directly focus on the modeling part of the lip motion generation, which is the contribution of this study.

\section{Methodology}
\label{sec:methodology}
\begin{figure}[t]
	\centering
	\includegraphics[trim=0mm 7cm 0cm 0mm, clip, width=\columnwidth]{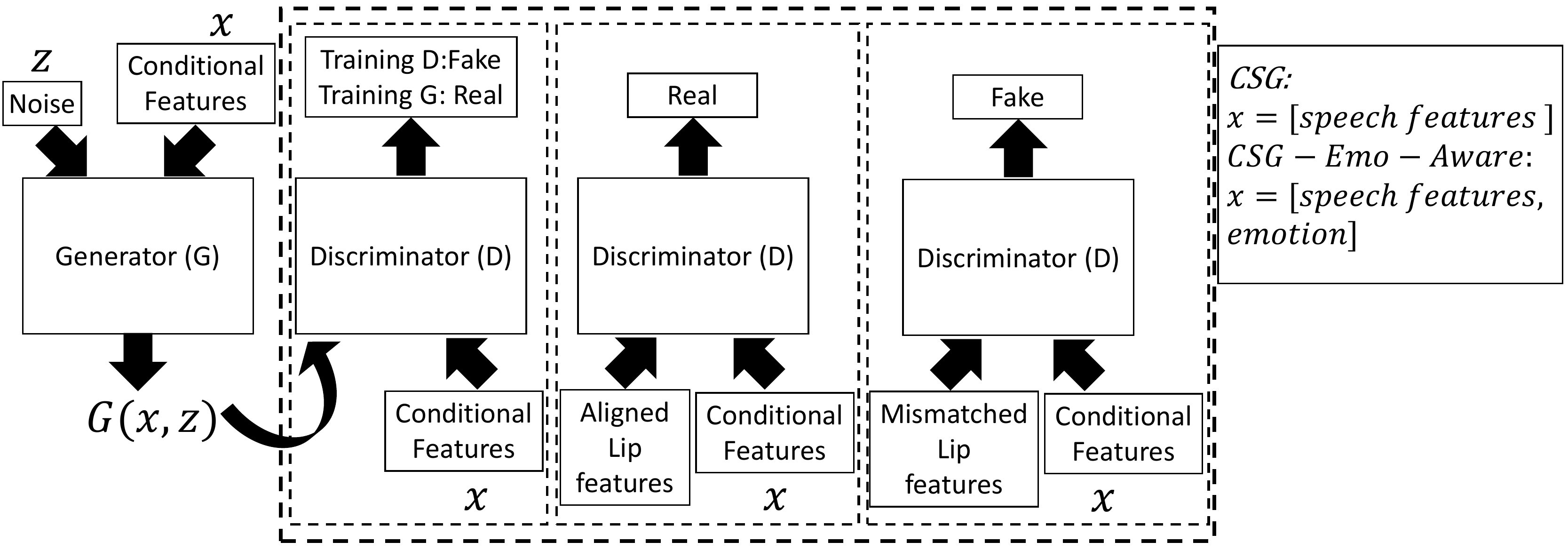}
	\caption{Proposed CSG model. Figure \ref{fig:CSGmodels} provides more detailed diagrams.}
	\label{fig:bigpic}
\end{figure}

The study proposes to generate lip motion driven by speech using adversarial training. The proposed framework corresponds to a conditional GAN for generating orofacial movements from audio features. Figure \ref{fig:bigpic} shows the overall framework for this model, which is called \emph{conditional sequential GAN} (CSG). The figure demonstrates how the generator and the discriminator are trained using the real and fake samples. The discriminator is trained to distinguish between the real and fake samples, where the real samples are the lip sequences aligned with the input audio, and the fake samples are either the lip sequences synthesized by the generator or the real lip samples which are not aligned with the input audio (i.e. mismatched). The generator is trained to fool the discriminator (i.e., the target label is real).
Two strengths of the approach are that (1) it does not require any lexical label, since it directly learns the mapping between speech and lip motion, and (2) it can be adapted to synthesize expressive behaviors when the intended emotion is provided as input. This section presents our proposed speech-driven framework for lip synthesis, describing the required building blocks and their roles in solving this problem.

\subsection{Bidirectional Long Short-Term Memory (BLSTM)}
\label{ssec:lstm}

We aim to predict orofacial movements from speech features. Since these sequences are time continuous signals, we  choose a model which can capture their temporal dependencies. \emph{Recurrent neural networks} (RNNs) use connection weights between consecutive hidden units to preserve temporal dependencies. However, training these models are probable to encounter  vanishing or exploding gradients \cite{Hochreiter_1997}. Extensions of RNNs such as LSTMs are proposed to efficiently handle this issue \cite{Hochreiter_1997}. We build our models based on (LSTMs). LSTMs associate memory cells with hidden units to keep track of past content. They use gating mechanisms to selectively manage the content stored in the memory cells. 

\begin{figure}
	\centering
	\includegraphics[trim = 0.5cm 4cm 10cm 0cm, clip, width=0.8\columnwidth]{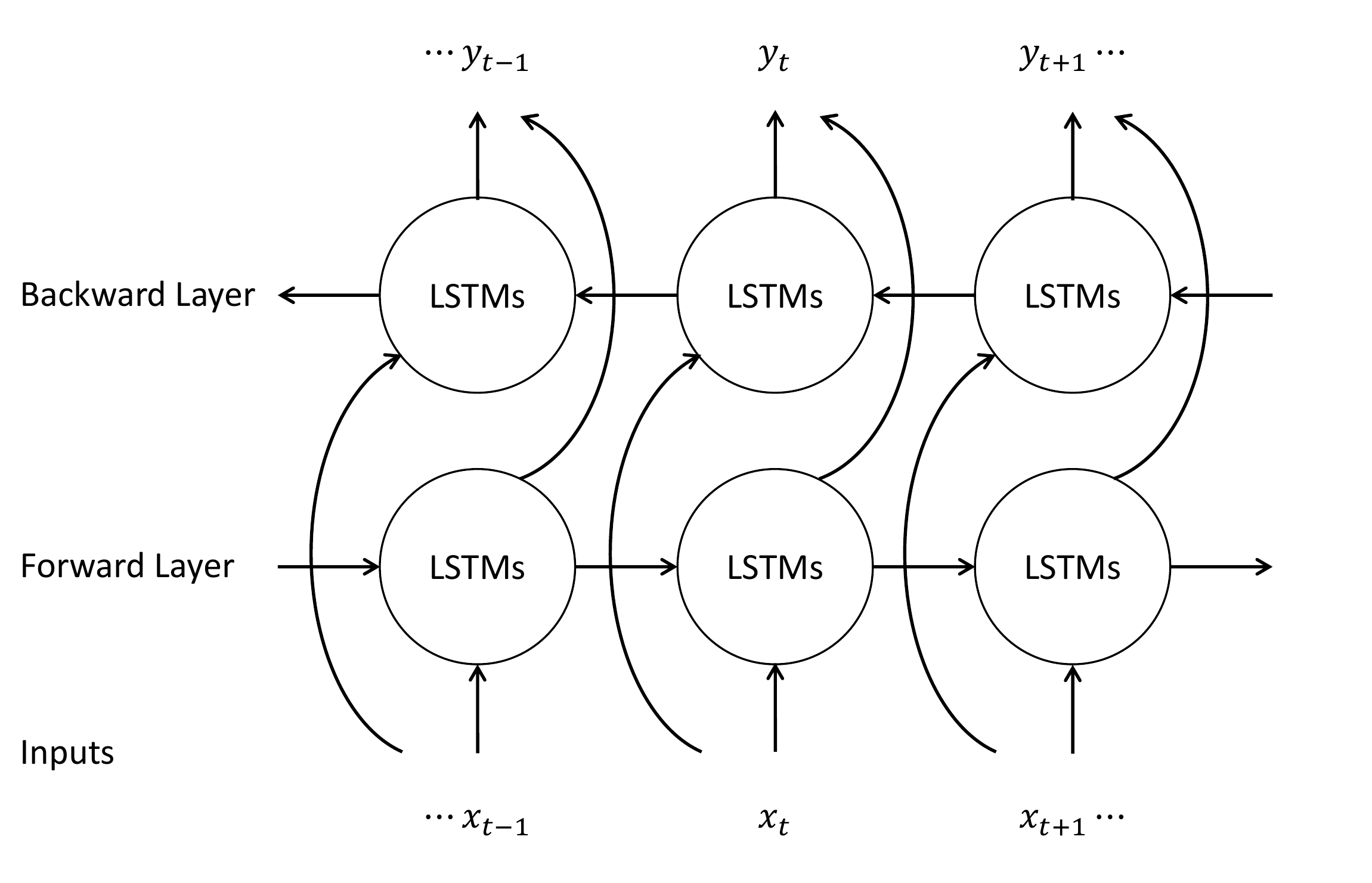}
	\caption{Illustration of BLSTM composed of forward and backward layers. The layer takes input $\mathbf{x}_t$ creating output $\mathbf{y}_t$}
	\label{fig:blstm}
\end{figure}

Incorporating future frames as well as the past frames can help the model to make better predictions. Therefore, our models are built with \emph{bidirectional LSTMs} (BLSTMs). These models consist of forward and backward paths of LSTMs, duplicating the number of hidden states (Fig. \ref{fig:blstm}). While this model can be used in real time using a short delay, we assume that we have the entire sequence of audio features, generating the sequences offline.

\subsection{Generative Adversarial Network (GAN)}
\label{ssec:GAN}

GANs are proposed as a generative model for learning the distribution of the data \cite{Goodfellow_2014}. The training in GAN is a mini-max game between two players, a generator (G) and a discriminator (D). The role of the discriminator is to distinguish between the samples generated by the generator (fake samples labeled as 0), and the samples from the original data (real samples labeled as 1). The role of the generator is to create samples given the input noise ($\mathbf{z}$), which resemble the real data, fooling the discriminator. This game can be achieved by the loss function in Equation \ref{eq:gan}, where $\mathcal{L}$ is the loss function, $E$ represents the expected value, $\mathbf{x}$ represents the real samples, $\mathbf{z}$ represents the input noise to the generator, $D(.)$ represents the discriminator function, and $G(.)$ represents the generator function.

\begin{equation}
\begin{split}
\underset{G}{\mathrm{min}}\ \underset{D}{\mathrm{max}}\ \mathcal{L}(D, G) =& E_{x\thicksim p_{data}(x)}\left[\mathrm{log} D(x)\right] \\
+ &E_{z\thicksim p_{z}(z)}\left[\mathrm{log} \left(1-D\left(G\left(z\right)\right)\right)\right]
\end{split}
\label{eq:gan}
\end{equation}

\subsection{Conditional Sequential GAN (CSG)}
\label{ssec:CSG}

\begin{figure}
	\centering
	\subfigure[CSG]{
		\includegraphics[trim = 2.cm 22.5cm 18.cm 26cm, clip, width=\columnwidth]{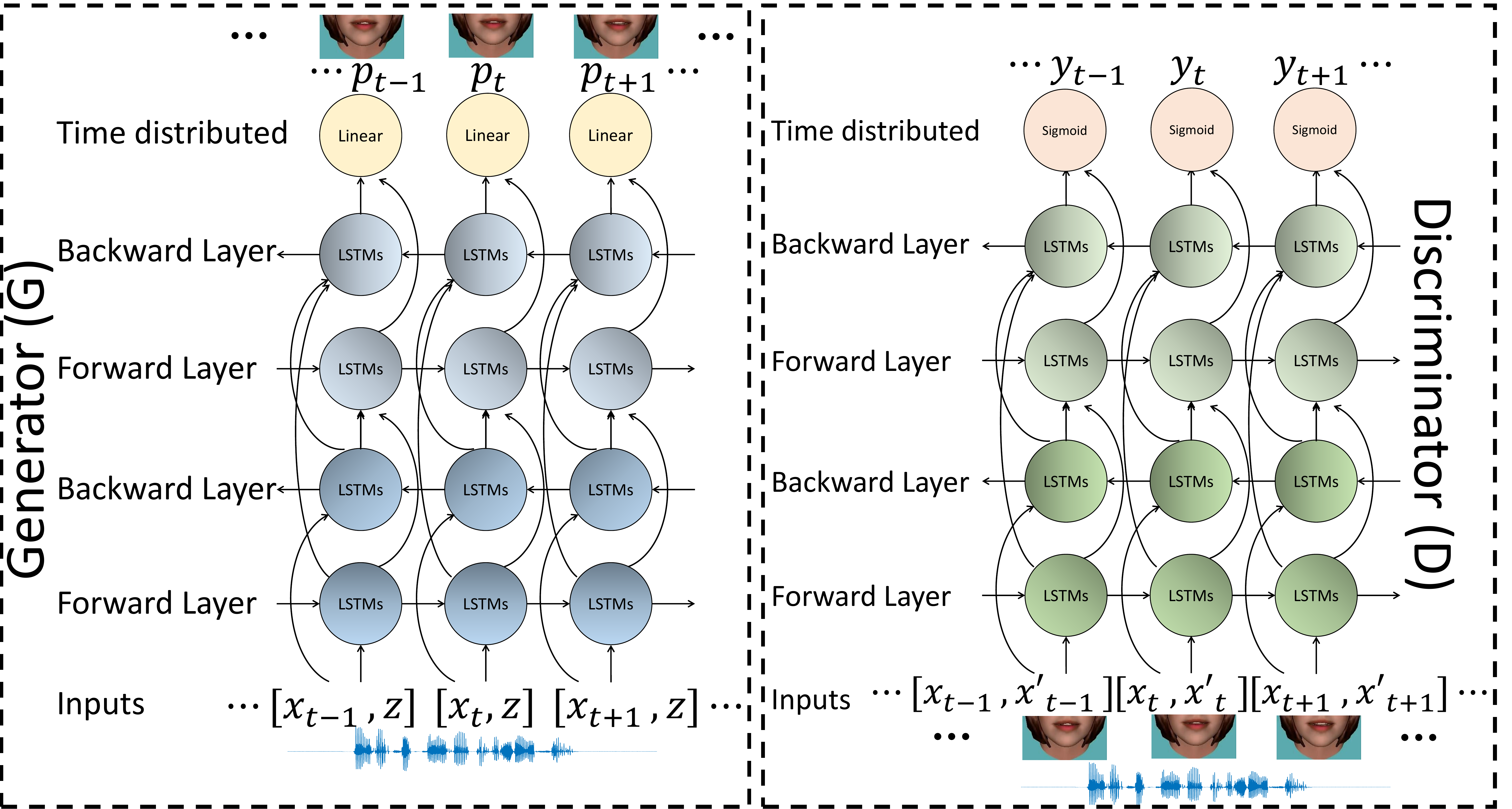}
		\label{fig:overview}
	}
	\subfigure[CSG-Emo-Adapted]{
		\includegraphics[trim = 1.7cm 22.5cm 6.cm 26cm, clip, width=\columnwidth]{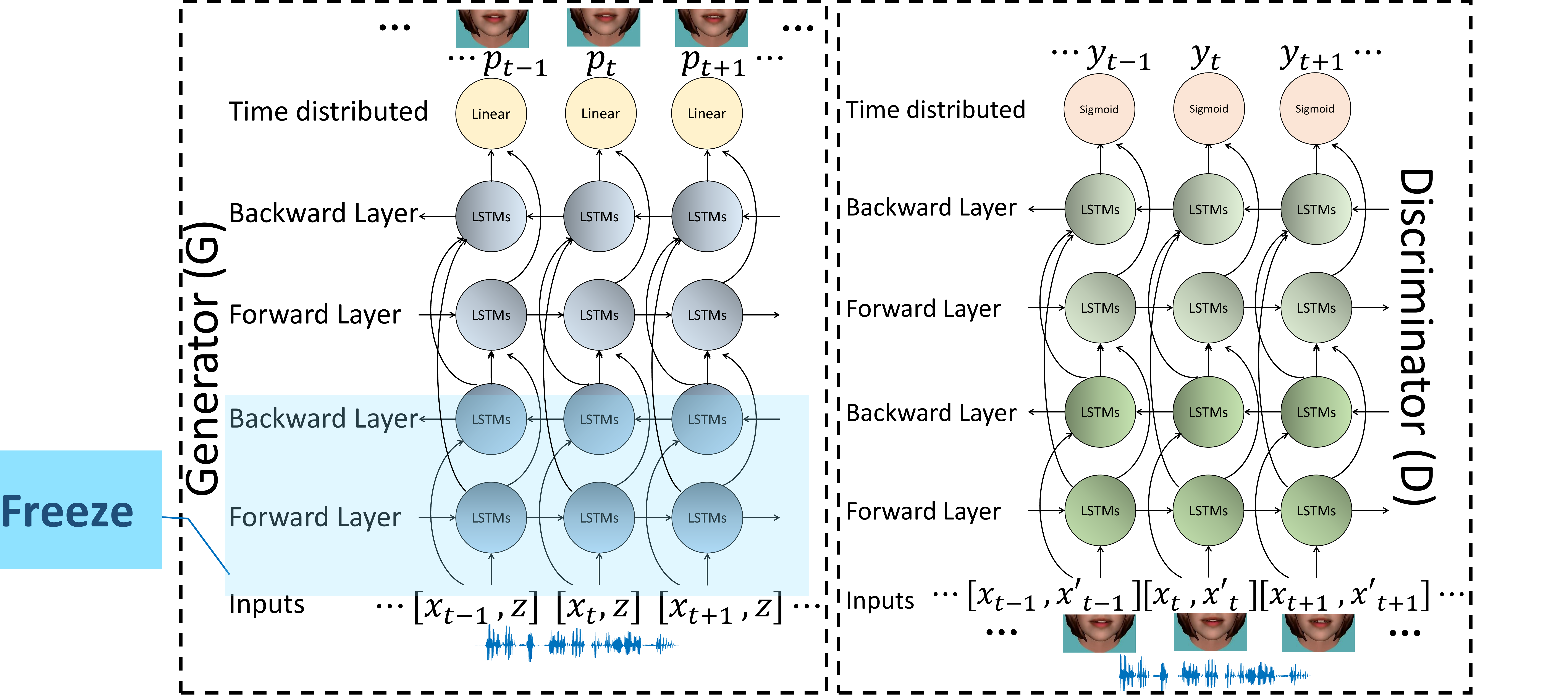}
		\label{fig:emoada}
	}
	\subfigure[CSG-Emo-Aware]{
		\includegraphics[trim = 2.cm 22.5cm 17cm 26cm, clip, width=\columnwidth]{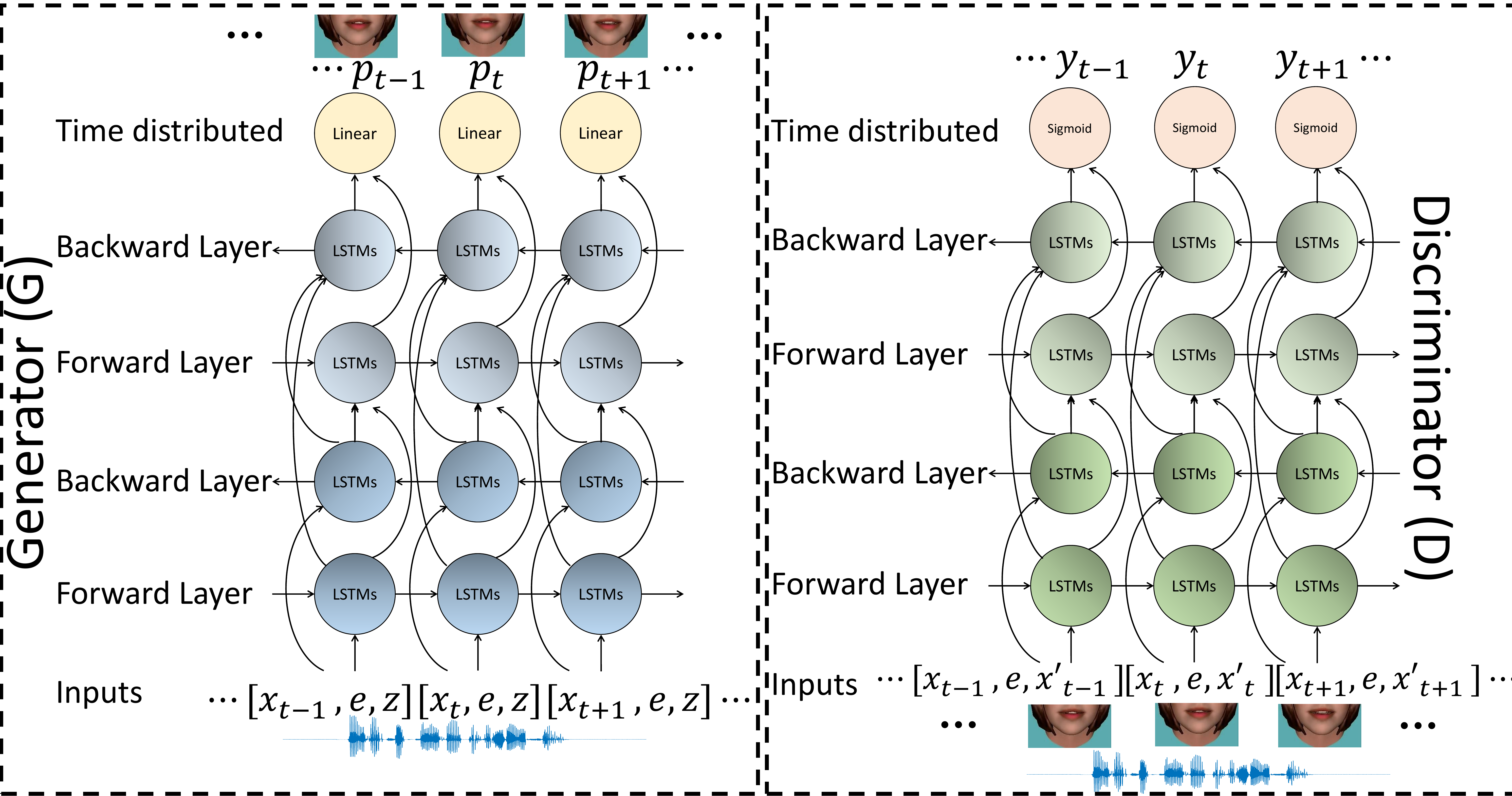}
		\label{fig:emocond}
	}
	\caption{The proposed frameworks to generate expressive lip motion sequences driven from speech, where $t$ represents the time frame index, $\mathbf{x}$ represents speech features, $\mathbf{p}$ represents the output of the generator, $\mathbf{x}'$ represents the orofacial pose, $\mathbf{y}$ represents the output of the discriminator, and $\mathbf{e}$ represents the vector with soft emotional labels.}		
	\label{fig:CSGmodels}
\end{figure}

Our proposed model is different from a simple GAN \cite{Goodfellow_2014}. We aim to drive the lip motion with acoustic features. Therefore, we propose to use a conditional GAN model, where the constraints to the discriminator and the generator are acoustic features (see Fig. \ref{fig:bigpic}). The input to our model is composed of a window of speech features plus random noise tied across the frames. The model maps the distribution of noise conditioned on the time-varying speech features to the distribution of original lip movements conditioned  on speech features. We call our model \emph{conditional sequential GAN} (CSG), which is shown in Figure \ref{fig:overview}. Previous studies have proposed different sequential GAN models to capture dynamics in videos \cite{Vondrick_2016,Tulyakov_2017}. However, previous conditional sequential GANs are implemented with static conditions tied across the input sequence \cite{Tulyakov_2017}. A key feature of our CSG model is that the input variable that conditions the GAN models is a time-varying signal (i.e., speech features).

Since we aim to learn the relationship between time-continuous signals (i.e., speech and lip movements), we build our cGANs with two layers of BLSTMs. We consider a linear output layer tied across all frames for the generator. We consider a sigmoid layer tied across all frames for the discriminator. We condition the generator and discriminator on the input features extracted from speech ($\mathbf{x}$) (Fig. \ref{fig:overview}).

Inspired by the matching-aware discriminator training strategy proposed by  \citet{Reed2016_2} for text-to-image synthesis, our learning strategy includes two kinds of fake samples during the training of the discriminator: samples generated by the generator, and original samples with lip motion and speech features extracted from different recordings. 
The first type of fake samples forces the generator to create realistic lip movements by decreasing the difference between the synthesized and actual lip trajectories. The second type of fake sequences forces the generator to capture the temporal relationship between lip motion and speech. By learning these two types of fake samples, the discriminator helps in training the generator to create lip motion sequences which are not only realistic, but also strongly coupled with the audio features. Although a conditional GAN model should theoretically learn these two types of fake samples by itself by using only samples from the generator (i.e., type one), using fake samples with uncoupled audio and lip motion emphasizes the importance of the temporal relationship between the modalities, which expedites the learning process (i.e., type two). 

The proposed CSG model needs to generate smooth trajectories for lip movements. Therefore, we use the same noise, $\mathbf{z}$, across all the input frames. It represents the global variations of conditional lip movements. To capture the dynamics of the movements, the CSG  relies on the time-varying speech features provided across the frames as evidence for the dynamics of the orofacial movements, which is captured by the LSTM units. The success of the sequential generator depends on two factors: each orofacial configuration generated at each frame needs to look realistic with respect to the speech features, and  the sequence generated by the generator needs to have realistic dynamics \cite{Tulyakov_2017}. Therefore, we use fake/real labels not only on the final frame, but also on all the intermediate frames from the discriminator. This approach allows us to minimize the loss function not only on the final frame of the sequence, but also on all the intermediate frames. Figure \ref{fig:overview} highlights that the discriminator considers the outputs across all the frames of the sequence. Our preliminary experiments demonstrated that this approach expedites learning. Note that we train our model by considering a fixed window length for both the generator and the discriminator.

\subsection{Expression-Aware CSG Models}
\label{sec:emotion-CSG}
The last building block in our proposed approach is to constrain the models on the target categorical emotion intended for the sentence, which is assumed to be an input for the model. We propose two expression-aware models as extensions of CSG, which utilize the categorical emotional labels during training and testing the models.

\subsubsection{Emotionally Adapted Conditional Sequential GAN (CSG-Emo-Adapted)}
\label{sssec:CSG-Emo-Adapted}
Figure \ref{fig:emoada} illustrates the first proposed approach. After training the CSG model with all the data, we separately adapt this model using the data associated with four emotions (i.e., the data with consensus labels for anger, happiness, sadness and frustration). \citet{Yosinski_2014} showed that the lower layers of DNNs are more generalizable than higher layers, which become more specific towards the primary task they are trained on. Therefore, we freeze the weights of the generator in the CSG on the first BLSTM layer, and fine tune the rest of the model, including the discriminator with the data associated with a given emotion. Freezing the weights is important to reduce the number of parameters to be learned given the reduced size of the data belonging to each emotion. We repeat this process for each emotion, creating emotion dependent models. The discriminator is the teacher of the generator. Therefore, it is important that the discriminator is fine-tuned with the adaptation data, so that the errors which correct the generator's mistakes are actually learned from the adaptation data. We hypothesize that this model helps the generator to synthesize more expressive lip motion sequences.

\subsubsection{Emotion-aware Conditional Sequential GAN (CSG-Emo-Aware)}
\label{sssec:CSG-Emo-Aware} 
Figure \ref{fig:emocond} illustrates the second proposed approach. This model conditions both the generator and the discriminator in the CSG model on the soft emotional labels of the speaking turn parametrized with the 6D vector explained in Section \ref{ssec:iemocap}. 
Compared with the CSG-Emo-Adapted model, this model better utilizes the IEMOCAP corpus, since all the segments are used, including the turns without consensus. The relationship between emotion and orofacial movements are assumed to be captured by the discriminator. We use real lip trajectories which are uncoupled with the acoustic and emotional features as fake samples. This approach helps the discriminator to learn this kind of fake instances, forcing the generator to create orofacial sequences that not only are coupled with speech, but also convey the emotional state of the speaker.

\section{Experimental Evaluation}
\label{sec:ExperimentalEvaluation}

\subsection{Implementation Details}
\label{ssec:Implementation}    
Our generator and the discriminator have two layers of BLSTMs. We set the number of nodes for the BLSTMs to 256 for the generator, and 128 for the discriminator. We implement our models using Keras with Theano as backend. We use \emph{adaptive moment estimation} (ADAM) \cite{Kingma_2014_2} as our optimizer. ADAM relies on the history of the gradient, in terms of its first and second moment, scaling the gradient to make the steps invariant to the gradient magnitude. This approach helps adapting the learning rate according to the changes in the loss at each iteration. For ADAM, we tried several learning rates $\left[0.001, 0.0001, 0.00001\right]$, selecting 0.0001, which gave the best loss reduction on the validation set. We set our batch size as 128 sequences with a fixed window size. We use the same window size as the one selected by \citet{Sadoughi_2018_2,Sadoughi_2018} which is 71 frames (591.7ms). This window size is also close to the the window size selected by \citet{Karras_2017}, which is 520ms.

We noticed that pre-training the generator is very helpful and expedites the training of the GAN.  The pre-training process relies on a DNN trained with BLSTM using \emph{concordance correlation coefficient} (CCC). This framework is the BLSTM-CCC baseline used in the experimental evaluation (Sec. \ref{ssec:baselineccc}). We pre-train the generator for 200 epochs. After the generator is pre-trained, we pre-train the discriminator by freezing the weights of the generator. We train the discriminator for 100 epochs.  After pre-training the models, we alternately train the generator and the discriminator on each batch. This scheme freezes the generator's weights, and updates the discriminator's weights on the current batch. Then, it freezes the discriminator's weights, and updates the generator's weights with the goal of fooling the discriminator. This goal is achieved by switching the labels of the fake samples when training the generator to fool the discriminator (i.e., switching the labels from 0 to 1). With this approach, the weights of the generator are updated with the objective of classifying the synthesized samples as real by the discriminator. We train all the CSG models for 50 epochs, alternating at each batch between updating the discriminator and updating the generator. All the adapted CSG models are also fine-tuned using this adversary scheme for 50 epochs.

The CSG models are pre-trained by maximizing the CCC. Equation \ref{eq:ccc} defines the CCC between two continuous variables $y$ (output)  and $t$ (target), where $\rho$ is the Pearson's  correlation coefficient between the two variables, $\mu_y$ and $\mu_t$ are the means of $y$ and $t$, and $\sigma_y^2$ and $\sigma_t^2$ represent the variances of $y$ and $t$. This loss function ($\ell$) favors high correlation between the predictions and the true values, while reducing the shift in the predicted values compared with the original ones. Optimizing this loss function not only reduces the \emph{mean squared error} (MSE), but also increases the Pearson correlation. It also increases the variance of the generated movements avoiding over smoothed trajectories.

\begin{equation}
\begin{split}
&CCC = \frac{2\rho\sigma_y\sigma_t}{\sigma_y^2+\sigma_t^2+(\mu_y-\mu_t)^2}\\
&\ell = 1-CCC
\end{split}
\label{eq:ccc}
\end{equation}

\subsection{Baselines}
\label{ssec:baseline}

We compare our model with three competitive baseline systems. Recent studies have shown that DNN-based approaches are more effective than HMM-based systems for this task \cite{Taylor_2016, Fan_2016,Parker_2017}. Therefore, we do not consider HMM-based solutions. 

\subsubsection{Sliding Window Deep Neural Network (SWDNN)}
\label{sssec:Baseswdnn}
\citet{Taylor_2016} proposed a model composed of three layers of \emph{rectified linear units} (ReLUs), with 2,000 nodes per layer, and a linear output layer to convert the input audio features  to orofacial movements over a smaller window centered at the middle frame of the input window. This model is trained to minimize the MSE between the predictions and real samples. During testing, the average of the predicted output frames are selected as the orofacial pose of the middle frame, moving one frame at a time to generate the entire sequence. We implemented the model following the description of the model, with the same input (340ms $\sim$41 frames) and output (100ms $\sim$13 frames) window sizes. Similarly, we use batch normalization on the layers to speed up the training, and we use a dropout $p=0.5$ on all the ReLU layers. We refer to this model as \emph{sliding window deep neural network} (SWDNN). 

\subsubsection{Bidirectional LSTM with MSE Objective (BLSTM-MSE)}
\label{ssec:baselinemse}

\citet{Fan_2016} proposed to use BLSTMs for learning the relationship between speech and orofacial movements, by minimizing the MSE between the predictions and the original movements. We implemented a model composed of two layers of 256 BLSTM units and a linear output layer, relying on the same objective (i.e., MSE). \citet{Fan_2016} implemented this model with varying length sequences. In our preliminary evaluation, we followed this approach by varying the length of the sequences, using the entire utterances. However, this approach generated over-smoothed trajectories that were not very appealing. Therefore, we train this framework with fixed window lengths, which generated more realistic sequences. We refer to this model as BLSTM-MSE.

\subsubsection{Bidirectional LSTM with Concordance Correlation Objective (BLSTM-CCC)}
\label{ssec:baselineccc}

This model is composed of two layers of 256 BLSTM units and a linear output layer (i.e., same as BLSTM-MSE). We define the loss function of this model based on CCC, inspired by the study of \citet{Sadoughi_2017}, which investigated facial movement prediction from speech. This model has the same lost function as our CSG models. This model is trained using a fixed window length. We refer to this model as BLSTM-CCC.

\subsection{Implementation Details for the Baselines}
\label{ssec:ImplBase}

We implement the baseline models using Keras with Theano as backend. The weights are initialized with the approach proposed by  \citet{Glorot_2010} ($W \sim U\left[-\frac{\sqrt{6}}{\sqrt{n_i+n_{i+1}}}, -\frac{\sqrt{6}}{\sqrt{n_i+n_{i+1}}}\right]$, where $U$ is the uniform distribution, $W$ is the weight between layers $i$ and $i+1$ and $n_{i}$ is the number of states for the $i^{th}$ layer).  We use ADAM as our optimizer, selecting a learning rate of 0.0001, since it gave the best loss reduction on the validation set for the baseline models. We set our batch size as 128 sequences with a fixed window size of 71 frames (591.7ms). We train all the baseline models for 200 epochs, except for the SWDNN model, which we train for an additional 800 epochs (the results on the validation set indicated that increasing the number of epochs reduced the error).

\subsection{Evaluation Metrics}
\label{ssec:Metrics}

The models are compared with objective and subjective evaluations. This section describes the metrics and procedure that are consistently used in the experimental evaluation.

\subsubsection{Objective Evaluation}
\label{ssec:objeval}
Objective evaluations of the results generated by GAN are usually provided by fitting a distribution to the generated samples, and getting the likelihood of the test samples in that distribution \cite{Goodfellow_2014}. This value shows how well the distribution of the generated samples matches the real samples. We use the Parzen window-based density estimation \cite{Goodfellow_2014}. Since we use conditional GANs, we provide the input features from the test set, and get the samples from the generator. To estimate the distributions, we treat each frame as one sample. To avoid the curse of dimensionality and increasing the error in the Parzen estimator, we use \emph{principal component analysis} (PCA) on the original samples to reduce the dimension of the samples from 45 to 15. A 15D vector preserves more than 95\% of the variance of the original orofacial data. We use cross validation to set the bandwidth of the Parzen estimator on the samples generated by the generator. We estimate the log-likelihood of the test samples (Sec. \ref{ssec:iemocap}) from the estimated distribution, reporting their average values and standard deviations. While we can always draw more samples from the CSG models by sampling different values from the noise distribution, we only generate one trajectory for each speaking turn, since the baseline systems can generate only one value per speech signal.

\subsubsection{Subjective Evaluation}
\label{ssec:subeval2}
\begin{figure}
	\centering
	\includegraphics[trim = 1cm 1cm 16cm 3cm, clip, width=0.7\columnwidth]{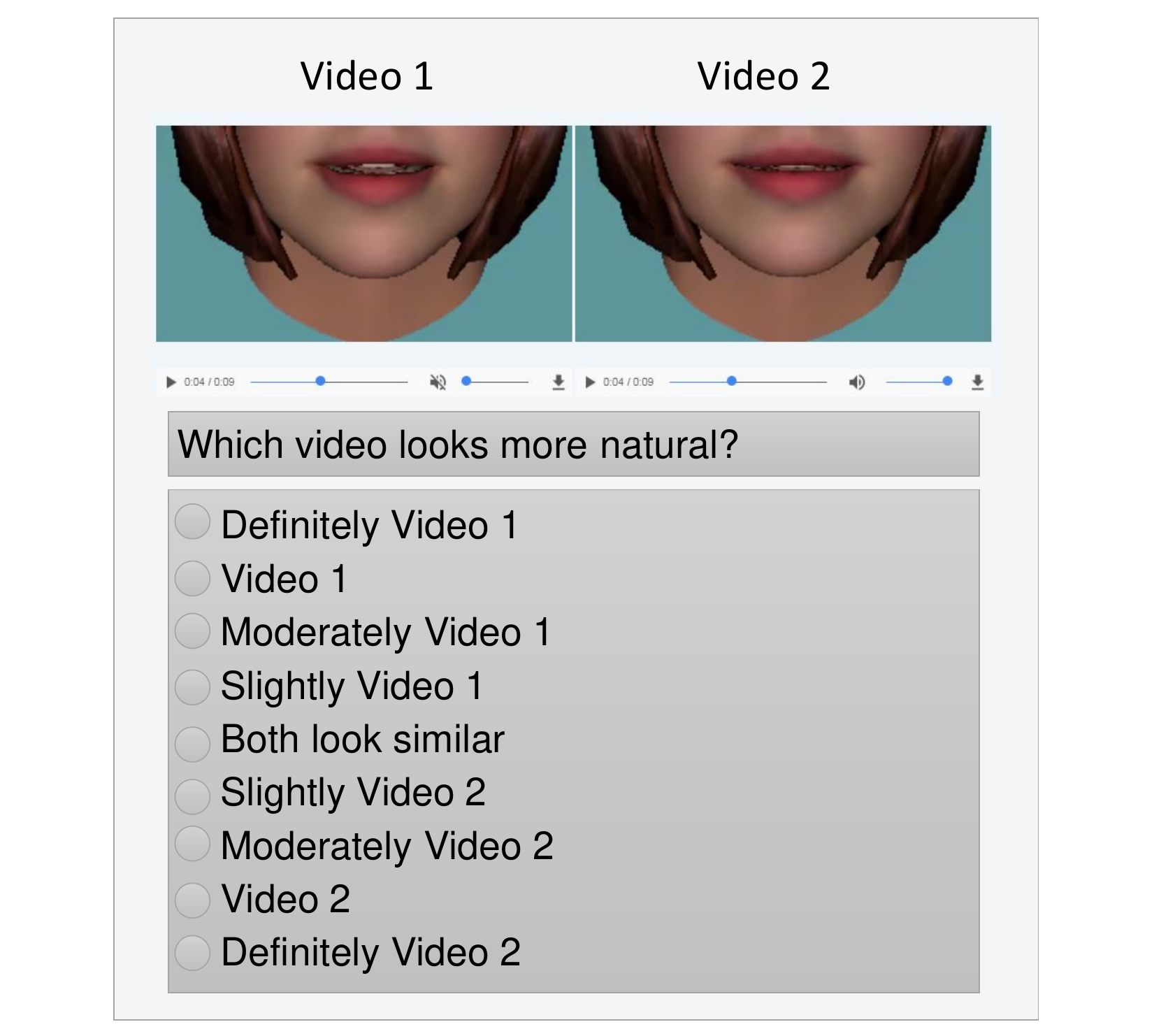}
	\caption{Interface used for our subjective evaluations using AMT.}
	\label{fig:mturk}
\end{figure}

The trajectories that we generated not only need to have similar distribution as the original sequences, but also  need to be perceived realistic. Therefore, we conducted subjective evaluations. People are more consistent in performing relative assessments than absolute ratings \cite{Yannakakis_2017}. Therefore, we perform subjective evaluations by asking for preference between two sequences generated with competing approaches. We ask ``which video looks more natural?'' in all the evaluations, except evaluations in Section \ref{ssec:emo-results}, where we additionally asked for the emotional perception elicited by the videos.  Figure \ref{fig:mturk} shows the interface. We provide multiple options to allow evaluators to convey their degree of certainty in the annotations ranging from ``\emph{definitely video 1}'' to ``\emph{definitely video 2}''. To report the results, we convert the selected options into percentage. For instance, the option ``\emph{definitely video 1}'' is mapped to 100\% for Video 1 and 0\% for Video 2, and the option ``\emph{moderately video 1}'' is mapped to 75\% for Video 1 and 25\% for Video 2.

We perform two different statistical tests on the comparison results. First, we compare the soft comparison assignments using a two way z-test with the null hypothesis that the two models being compared are perceived as similar (i.e., $h_0:$ MEAN $=50\%$). Second, we convert the soft assignment labels into hard assignments, by using Equation \ref{eq:prop}, where $i$ represents the $i^{th}$ sample and $n$ is the total number of samples.  For ties (i.e., 50\%-50\%), we assign one vote to each video. This approach allows us to compare the two models using a statistical proportion test on the hard assignments.

\begin{equation}
p = \frac{\sum_{i=1}^{i=n} 1 \left( e_i \geqslant 50\right)}{n+\sum_{i=1}^{i=n} 1 \left( e_i = 50\right)}
\label{eq:prop}
\end{equation}

All our subjective evaluations are conducted on \emph{Amazon mechanical turk} (AMT). For subjective evaluations, we randomly select five turns per emotion (i.e., 20 videos in total), and rendered their videos using the trajectory generated by the models. We consider the three baseline models (i.e., SWDNN, BLSTM-MSE and BLSTM-CCC), the three proposed CSG methods (i.e., CSG, CSG-Emo-Adapted, and CSG-Emo-Aware), and the original trajectories from the motion capture data. Therefore, we have 20 videos for each of the seven conditions. We also render 10 videos per emotional class as explained in Section \ref{ssec:emo-results}.

The evaluators compare two videos at a time created for the same sentence. The placement of the videos and the ordering of the pairs are randomized throughout each task. They complete 20 comparisons per \emph{human intelligent tasks} (HITs). The  question is shown after the annotator plays the two videos, reducing the chance of evaluators answering the questions before watching the videos. We use the Cronbach's alpha to quantify the agreement between evaluators. We limits the pool of annotators to people who have performed well in our previous crowd-sourcing evaluations \cite{Burmania_2016_2, Sadoughi_2017_3, Lotfian_201x}.

\section{Results}
\label{sec:results}

\subsection{Noise Dimension}
\label{ssec:noiseDimension}

An important parameter of our model is the dimension of the noise. We use an $m$-dimensional Gaussian noise with diagonal covariance matrix and zero mean. To choose the dimension of the noise, we used the CSG model, changing $m\in \{1, 10, 40, 80, 150\}$. We performed subjective evaluations on 10 videos generated by each model from the validation set. Each video is compared with the video rendered with the original lip motion sequences. The results provide indirect comparisons between the models with different noise dimensions. We use the protocol described in Section \ref{ssec:subeval2} for AMT. 

\begin{figure}
	\centering
	\includegraphics[trim =0cm 0.0cm 0cm 0cm, clip, width=0.9\columnwidth]{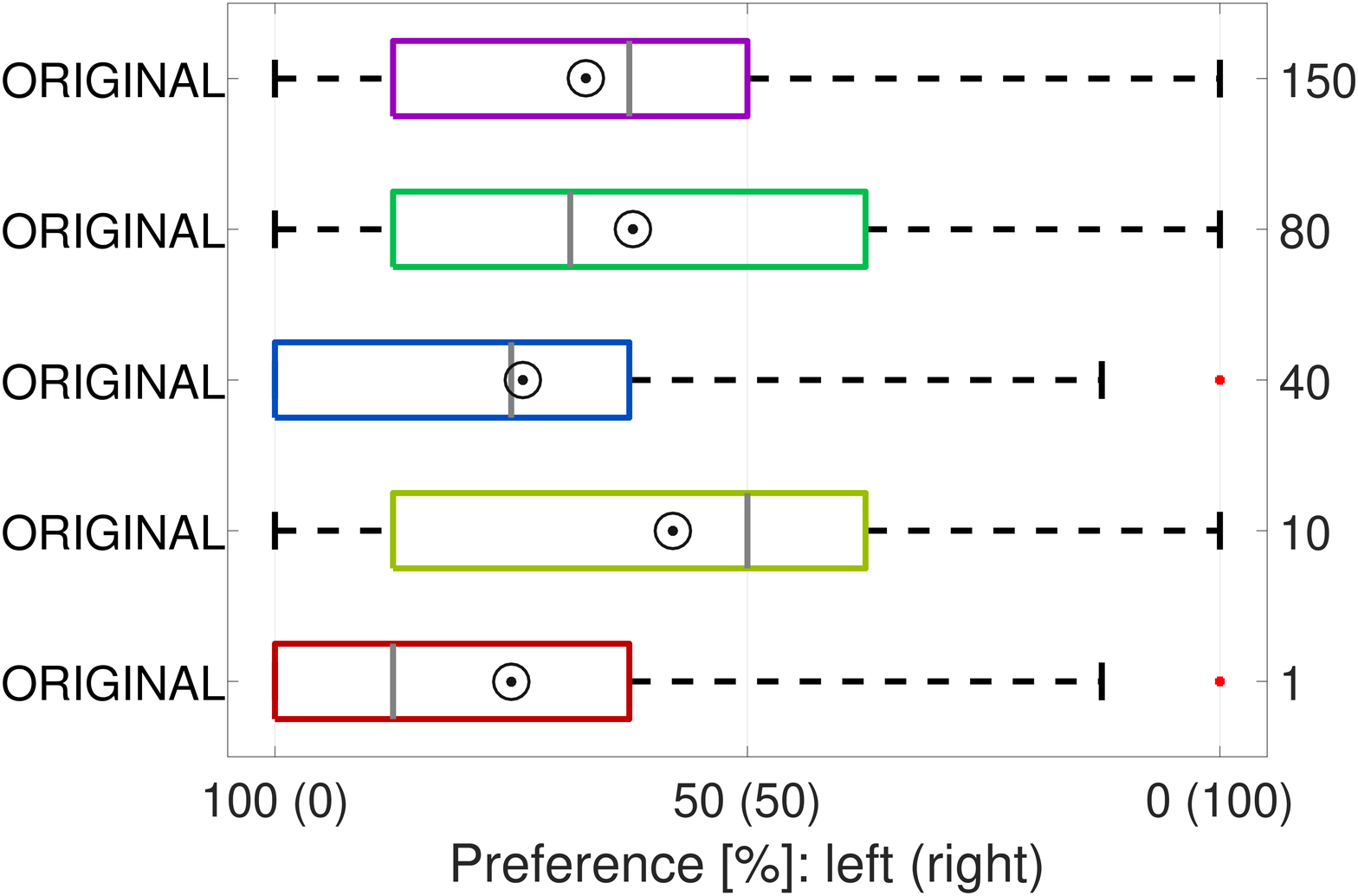}
	\caption{Comparison of the CSG models for different dimensions of the noise. The bars represent the first and third quartiles. The circle represents the mean values for each condition, the dash lines represent the minimum and maximum values, the vertical gray lines represent the medians, and the red dots represent outliers.}
	\label{fig:ndselect}
\end{figure}

We recruited 15 evaluators for this evaluation, each comparing 10 pairs of videos, resulting in three evaluations per video. Figure \ref{fig:ndselect} gives the results. The Cronbach's alpha between the annotators are $\alpha_{1}=0.72$, $\alpha_{10}=0.65$, $\alpha_{40}=0.78$, $\alpha_{80}=0.50$ and $\alpha_{150}=0.73$ (the subscript of $\alpha$ indicates the noise dimension). We discarded the evaluations of two raters whose average pairwise Cronbach's alpha was less than zero, repeating the HIT with other raters. As expected, the average of the preferences are shifted towards the original sequences. The results suggest that $m=10$ and $m=80$ are the most competitive models (i.e., the bars are shifted toward the center). We select $m=10$ as the dimension for the noise for the rest of the experimental evaluation. 

\subsection{Comparing the CSG Model with the Baselines}
\label{ssec:csgbase}

This section compares the CSG model with the three baseline models: the SWDNN, BLSTM-MSE, and BLSTM-CCC approaches. We compare these models with objective and subjective evaluations.

\noindent
\textbf{Objective Evaluation:}
We estimate the distribution of synthesized samples generated by the CSG and baseline models using the Parzen window density estimator. We generate 555K samples per model. Table \ref{tab:ll} gives the mean and standard deviations of the log-likelihood of the test samples in the fitted distributions. The proposed CSG model is significantly better than all other alternative baselines. Note that all the pairwise comparisons in this table are statistically different (z-test: $p$-value $<0.0001$). The results demonstrate higher log-likelihoods for the CSG model compared with the baselines. Interestedly, BLSTM-CCC outperforms BLSTM-MSE showing the benefit of using CCC as a cost function.

\begin{table}[t]
	\centering
	\caption{Comparing the results generated with the CSG and the baseline models in term of log-likelihood of the test samples in the estimated distribution by the Parzen estimator. All the pairwise comparison are statistically significant ($p$-value $<0.0001$)}
	\begin{tabular}{ c| c}
		\hline
		\multirow{2}{*}{model} & log-likelihood\\
		& MEAN (STD)\\
		\hline\hline
		SWDNN & -207.412 (268.452) \\
		BLSTM-MSE & -190.642 (318.110) \\
		BLSTM-CCC & -143.234	(317.674) \\
		CSG & \textbf{-125.797 (241.979)}\\
		\hline
	\end{tabular}
	\label{tab:ll}
\end{table}

\begin{figure}
	\centering
	\includegraphics[trim =0cm 0.0cm 0cm 0cm, clip, width=0.9\columnwidth]{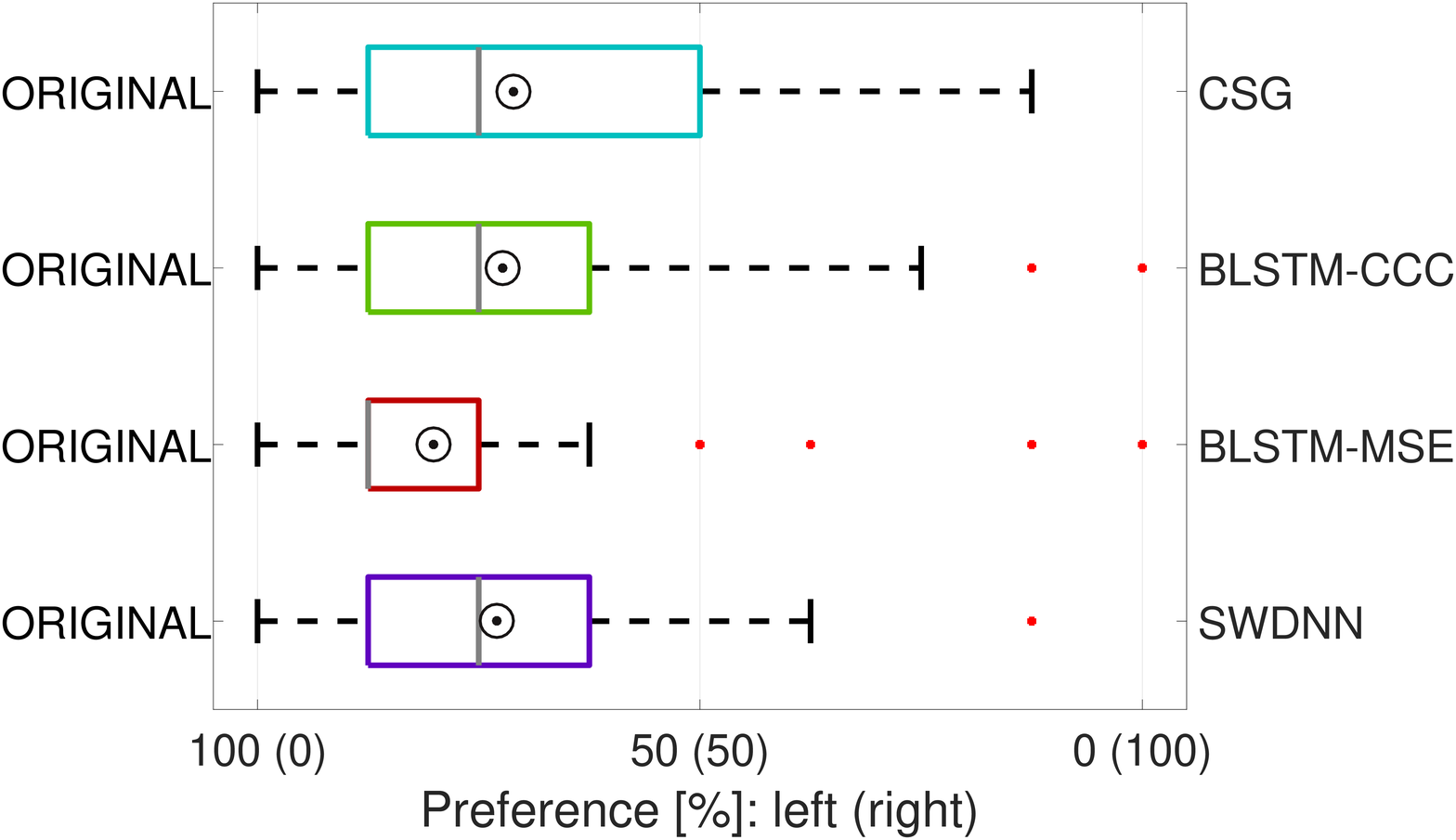}
	\caption{Comparison of the CSG model and the baseline models with videos generated with the original lip motion sequences. The bars represent the first and third quartiles. The circle represents the mean values for each condition, the dash lines represent the minimum and maximum values, the vertical gray lines represent the medians, and the red dots represent outliers.}
	\label{fig:subeval}
\end{figure}

\noindent
\textbf{Subjective Evaluation:}
The first phase of the subjective evaluations compares the animation synthesized by each of the models (CSG, SWDNN, BLSTM-MSE and BLSTM-CCC) with videos generated with the original motion capture recordings. We recruited 16 evaluators who annotated 20 pairs of videos, resulting in four evaluators per comparison. Figure \ref{fig:subeval} shows the result of these comparisons, where the agreement between evaluators in terms of the Cronbach's alpha are $\alpha_\mathit{SWDNN}=$0.88$, \alpha_\mathit{BLSTM-MSE}=$0.91 and   $\alpha_\mathit{BLSTM-CCC}=$0.83,  and $\alpha_\mathit{CSG}=$0.82. The z-test shows that the means of all these ratings are not equal to 50\% ($p$-value $<1e^{-12}$), indicating that the original sequences are preferred. Notice that these approaches are speech-driven models that do not rely on transcriptions, so the synchronization of the lips is not perfect. Therefore, it is expected that videos generated with original trajectories will be preferred by the evaluators. We estimate the proportion of preference for the original motion capture data with each of these models using Equation \ref{eq:prop}. The original sequences are preferred 87\% over the SWDNN model, 92\% over the BLSTM-MSE model, 78\% over BLSTM-CCC model, and 76\% over the CSG model. While the annotators preferred the videos with the original sequences (all these proportions are statistically significant --$p$-value $<0.01$), the CSG and BLSTM-CCC models are the approaches where the preferences are closer to 50\%. 

\begin{figure}
	\centering
	\includegraphics[trim =0cm 0.0cm 0cm 0cm, clip, width=0.9\columnwidth]{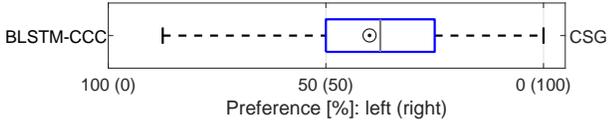}
	\caption{Comparison of the CSG model with the baseline (BLSTM-CCC). The figure follows the same convention used in Figure \ref{fig:subeval}.}
	\label{fig:subevalcsgbase}
\end{figure}

Figure \ref{fig:subeval} provides indirect comparisons between the models. The second phase of the subjective evaluations directly compares our proposed CSG model with the BLSTM-CCC model, which was the most competitive baseline model in the indirect comparisons. We use 20 videos synthesized by the CSG and BLSTM-CCC approaches. We recruit four raters for this task, who evaluated the 20 videos using the approach described in Section \ref{ssec:subeval2} (four evaluations per comparison). Figure \ref{fig:subevalcsgbase} shows the results, where the Cronbach's alpha between the annotators is $\alpha=$0.61. The z-test shows that the mean of the soft preferences is not equal to 50\% ($p$-value $<1e-5$), which indicates a clear preference for the proposed CSG model. Using Equation \ref{eq:prop}, we observe that 68\% of the evaluators preferred the CSG model over the BLSTM-CCC model. The trend is statistically significant according to the proportion test ($p$-value $ < 0.01$). 

Objective and subjective evaluations clearly show better performance for the CSG model over the baseline methods. The next section evaluates the benefits introduced by considering emotion in the expression-aware CSG models. 

\begin{table*}[t]
	\centering
	\caption{Comparing the CSG model with the two expression-aware CSG models for each emotional category. The values correspond to the log-likelihood of the test samples in the estimated distribution by the Parzen estimator. Asterisks indicate when the expression-aware CSG models are significantly better than the CSG model (*: $p$-value $<0.05$, **: $p$-value $<0.01$, ***: $p$-value $<0.001$).}
	\begin{tabular}{ c| c c c c }
		\hline
		\multirow{2}{*}{model}& \multicolumn{4}{c}{log-likelihood: MEAN (STD)}\\
		\cline{2-5}
		& ang & hap & sad & fru\\
		\hline\hline
		CSG &-76.410	(130.751)& -77.139	(183.269)&\textbf{-162.832	(208.430)}& -148.145	(239.602)\\
		CSG-Emo-Adapted &\textbf{-72.495	(119.353)}*** &\textbf{-75.365	(168.302)*}& -163.679	(229.455)&\textbf{-136.634	(226.355)}***\\
		CSG-Emo-Aware & -74.201	(116.977)**&-81.103	(189.359)***&-179.041	(245.666)***&-147.291	(232.307)\\
		\hline
	\end{tabular}
	\label{tab:ll-emo}
\end{table*}

\subsection{Expression-Aware CSG Models}
\label{ssec:emo-results}

This section evaluates the CSG-Emo-Aware and CSG-Emo-Adapted models. The objective evaluations consider the log-likelihood estimations (Sec. \ref{ssec:objeval}), and the accuracy of emotion classifier trained on the original data and tested on the synthesized results. The subjective evaluations consider the preference and expressiveness of the expression-aware models. 

\noindent
\textbf{Objective Evaluation:}
We estimate the distribution of the samples using the Parzen window density estimator. The number of the test samples (i.e., frames) across emotional classes are 62K for anger, 107K for happiness, 92K for sadness and 106K for frustration. We generate the same number of samples using each of the models. Table \ref{tab:ll-emo} gives the log-likelihood of the test samples evaluated on the distribution of the generated samples. All the pairwise comparisons between the CSG model and each of the expression-aware CSG models are statistically significant (z-test: $p$-value $<0.05$), with two exceptions: the comparison between CSG and CSG-Emo-Adapted for sadness and the comparison between the CSG and CSG-Emo-Aware for frustration. These results indicate that adding emotion in the models help in generating samples that are closer to the original sequences. Table \ref{tab:ll-emo} shows that the CSG-Emo-Adapted model constantly achieves better results than the CSG-Emo-Aware model. All the pairwise comparisons between the CSG-Emo-Aware and CSG-Emo-Adapted models are statistically significant ($p$-value $<0.05$). Adapting the top layers is an effective method to create expressive-aware models for lip motion.

We evaluate whether the generated lip movements convey emotional cues by training an emotion classifier. Using the same train, test and validation partitions used for the models, we train  a categorical emotion classifier on the original motion capture data. The classification tasks use lip motion sequence to recognize anger, happiness, sadness, and frustration. Since the emotional labels are assigned to each speaking turn, we extract sentence-level features by extracting statistics from the 45D orofacial pose parameters. The statistical features include mean, median, first quartile, third quartile, minimum, maximum and standard deviation, resulting in a 315D feature vector (45 parameters $\times$ 7 functionals).  We have 1,898 training samples across anger (359), happiness (525), sadness (390) and frustration (624). The validation set has 624 speaking turns (anger-132, happiness-187, sadness-111, and frustration-194), and the test set has 617 speaking turns (anger-129, happiness-169, sadness-152, and frustration-167). We train a SVM classifier, maximizing the F$_1$-score (i.e., F$_1$-score$=2\times\frac{precision\times recall}{precision+recall}$) on the validation set to determine the kernel function and the soft margin parameter. The best result on the validation set was obtained with a linear kernel and a soft margin equals to $c=0.8$. We evaluate this model on the test set, using the original motion capture recordings and the lip trajectories generated by the CSG models. Table \ref{tab:emo_acc} shows the results in terms of accuracy, recall, precision, and F$_1$-score. The classifier tested with the original data achieves an F$_1$-score of 61.5\%. The same classifier tested with the samples generated by the CSG model achieves an F$_1$-score of 37.7\%. These results show that the CSG model does not preserve the emotional cues in the lip motion trajectories. This problem is overcome by the expression-aware CSG models. The same classifiers tested with the samples generated by the CSG-Emo-Aware and CSG-Emo-Adapted models achieve F$_1$-scores of 62.5\% and 70.8\%, respectively. These results are statistically significantly better than the F$_1$-score achieved when using samples generated by the CSG model ($p$-value $<1e^{-8}$).  Even though we train with the original data and test with synthesized data, the emotion classifiers are able to recognize emotions with similar or better accuracy than when we test with the actual lip motion trajectories. These results demonstrate that the proposed expression-aware CSG models generate lip motion trajectories conveying expressive cues similar to the original recordings.

\begin{table}
	\centering
	\caption{Emotion recognition results over the synthesized orofacial movements by the models in terms of accuracy (Acc.), precision (Prec.), recall (Rec.) and F$_1$-score.}
	\begin{tabular}{l |c c c c}
		\hline
		\multirow{1}{*}{Orofacial Source} 		&Acc. & Prec. & Rec. & F$_1$-score\\
		&[\%]&[\%]&[\%]&[\%]\\
		\hline\hline
		Original &62.6&66.6&60.9&61.5 \\
		\hline
		CSG 	 &39.9&45.8&38.4&37.7\\
		CSG-Emo-Adapted&\textbf{70.5}&\textbf{74.8}&\textbf{69.8}&\textbf{70.8}\\
		CSG-Emo-Aware&62.2&63.9&61.8&62.5 \\
		
		\hline
	\end{tabular}
	\label{tab:emo_acc}
\end{table}

\begin{figure}
	\centering
	\includegraphics[trim =0cm 0.0cm 0cm 0cm, clip, width=0.9\columnwidth]{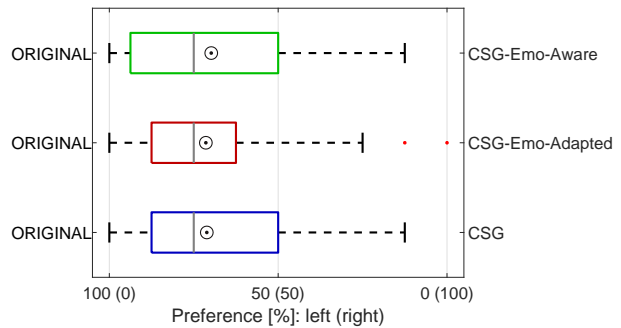}
	\caption{Comparison of the CSG model and expression-aware CSG models with videos generated using the original lip motion sequences. The figure follows the same convention used in Figure \ref{fig:subeval}.}
	\label{fig:subevale}
\end{figure}

\noindent
\textbf{Subjective Evaluation:}
We also perform subjective evaluations on the results, starting with indirect comparisons where the original sequences are used as reference. We recruiting eight evaluators for the expression-aware models, who were asked to evaluate 20 pairs of videos (four evaluators per task). Figure \ref{fig:subevale} shows the results, where we repeat the results obtained for the CSG model presented in Figure \ref{fig:subeval}. The agreements between the evaluators in terms of Cronbach's alpha are $\alpha_\mathit{CSG}=$0.82, $\alpha_\mathit{CSG-Emo-Adapted}=$0.79, and $\alpha_\mathit{CSG-Emo-Aware}=$0.85. The z-test shows that the evaluators prefer the original models, as expected ($p$-value $<1e^{-14}$). Using Equation \ref{eq:prop}, the proportions of preferences for the original motion capture data with each of the models are 76\% over the CSG model, 80\% over the CSG-Emo-Adapted model and 70\% over the CSG-Emo-Aware model. All these proportions are statistically greater than 50\% ($p$-value $<0.01$), which is not surprising.

\begin{figure}
	\centering
	\includegraphics[trim =0cm 1.7cm 0cm 0cm, clip, width=0.9\columnwidth]{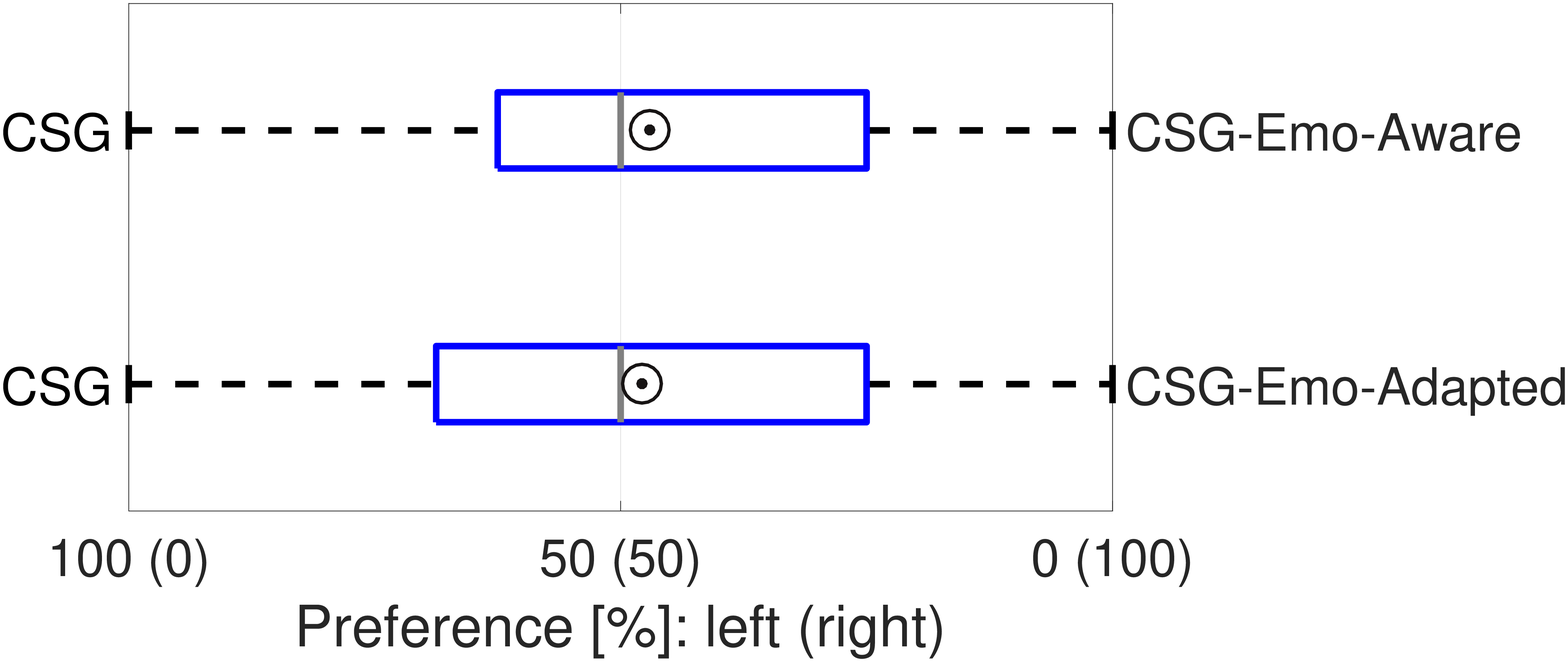}
	\caption{Comparison of the CSG model with the two expression-aware models. The figure follows the same convention used in Figure \ref{fig:subeval}.}
	\label{fig:subevalgemo}
\end{figure}

\begin{figure}
	\centering
	\includegraphics[trim =0cm 0.0cm 0cm 0cm, clip, width=0.9\columnwidth]{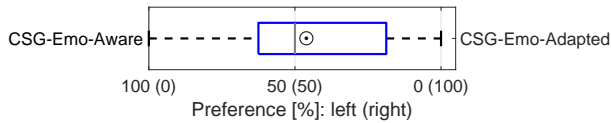}
	\caption{Comparison of the two expression-aware CSG model. The figure follows the same convention used in Figure \ref{fig:subeval}.}
	\label{fig:subevalemo}
\end{figure}

We also perform subjective evaluations to directly compare two videos where one of the lip motions were generated by the CSG model and the other with the expression-aware models. We generate 20 videos for the CSG-Emo-Adapted model and 20 videos for the CSG-Emo-Aware model, creating 40 video pairs. We recruit eight evaluators who compare 20 pairs of videos, resulting in four evaluations per comparison. Figure \ref{fig:subevalgemo} gives the results of this evaluation. The Cronbach's alpha between the annotators are $\alpha_\mathit{CSG-Emo-Adapted}=$0.48 and $\alpha_\mathit{CSG-Emo-Aware}=$0.63. While the evaluators slightly prefer the expressive-aware models (i.e., bars are closer to the expressive-aware models), the preference are not statistically significant (z-test, $p$-value $=0.48$ for CSG-Emo-Adapted, and $p$-value $=0.29$ for CSG-Emo-Aware). We directly compare the CSG-Emo-Aware and CSG-Emo-Adapted models. We recruited four evaluators to compare the 20 videos generated by these models,  resulting in four evaluations per comparison. The Cronbach's alpha between the annotators is $\alpha=$0.70. Figure \ref{fig:subevalemo} shows the results which show a small preference for the CSG-Emo-Adapted model, although the preference is not statistically significant (z-test, $p$-value $=0.27$). We estimate the proportion preference using Equation \ref{eq:prop}, which shows that 55\% of the evaluations prefer the CSG-Emo-Adapted model. The proportion test shows that the preference is not statistically significant ($p$-value $=0.25$).

\begin{figure}
	\centering
	\subfigure[ang: ``Which looks angrier?'']{
		\includegraphics[width=\columnwidth]{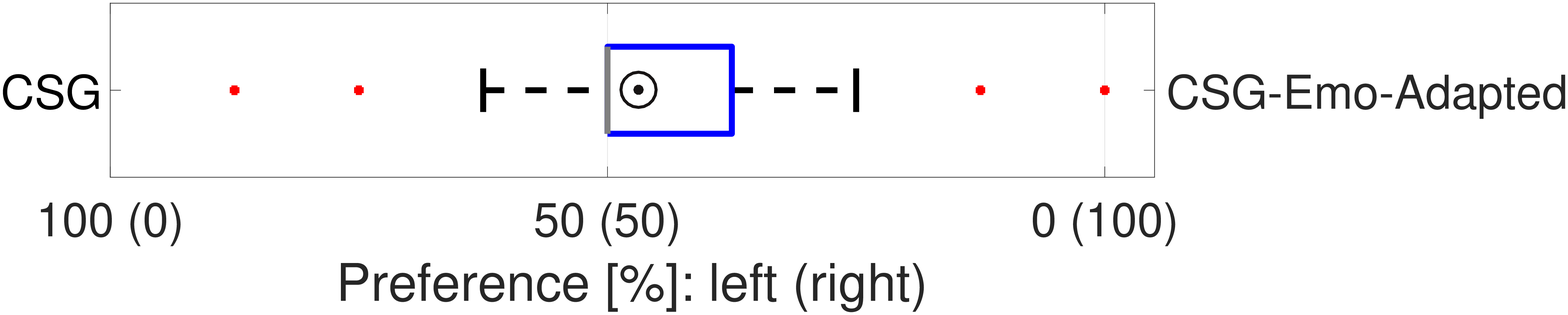}
	}
	\subfigure[hap: ``Which looks happier?'']{
		\includegraphics[width=\columnwidth]{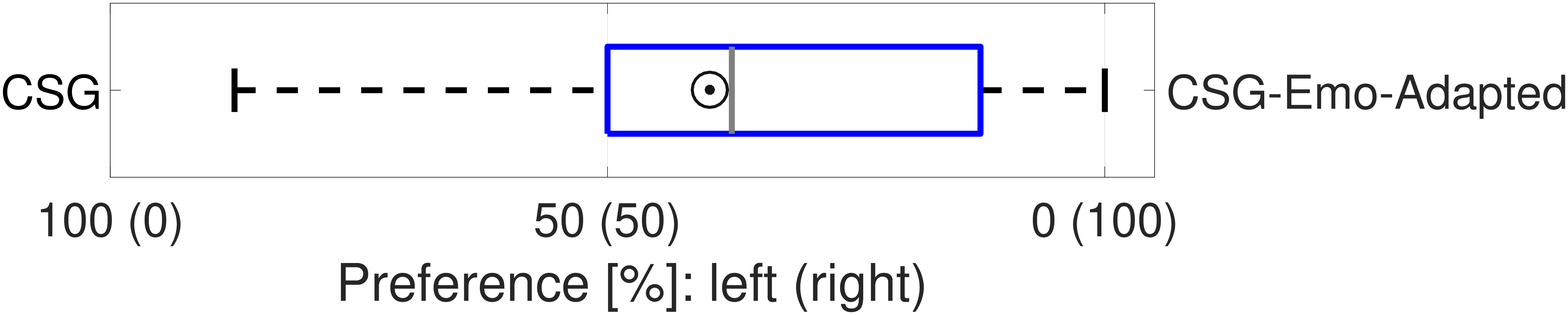}
	}
	\subfigure[sad: ``Which looks sadder?'']{
		\includegraphics[width=\columnwidth]{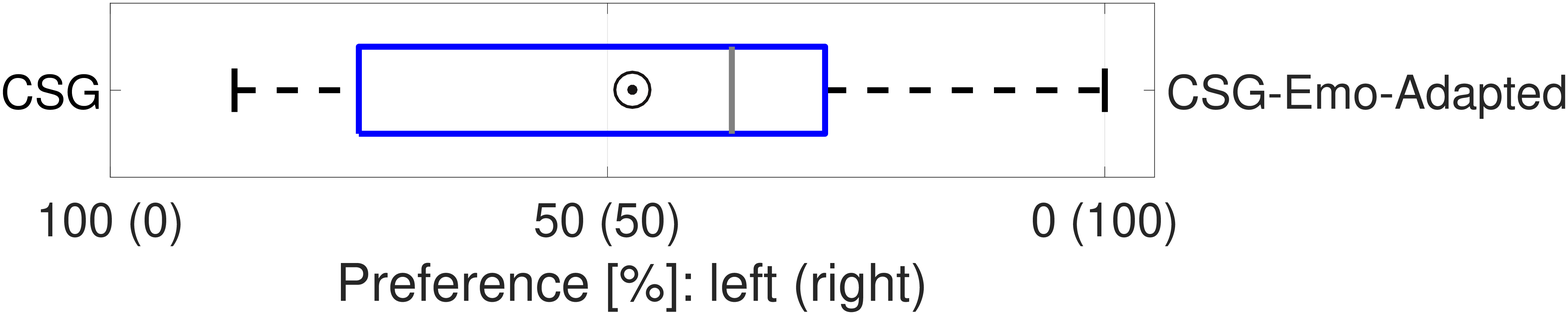}
	}
	\subfigure[fru: ``Which looks more frustrated?'']{
		\includegraphics[width=\columnwidth]{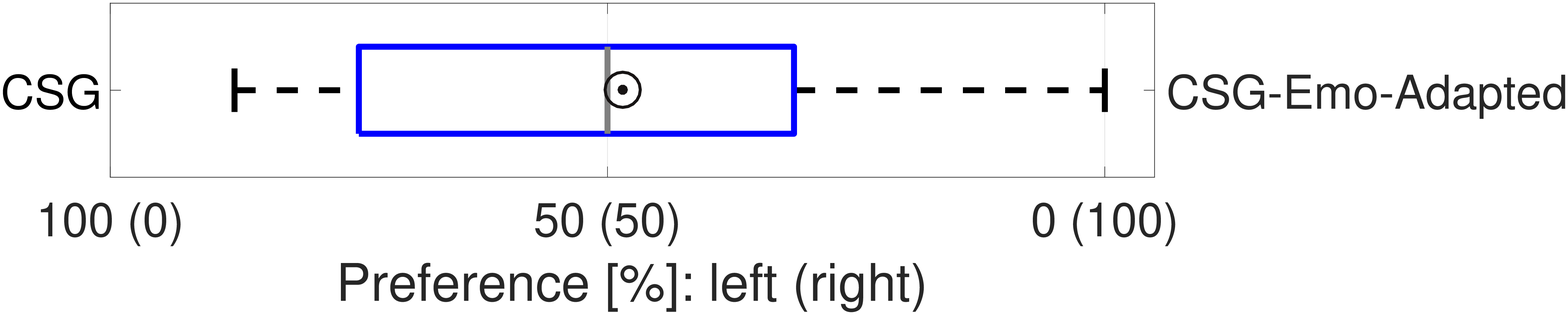}
	}
	\caption{Comparison of the perceived target emotional category elicited by the CSG and CSG-Emo-Adapted models. The figure follows the same convention used in Figure \ref{fig:subeval}.}
	\label{fig:subemo}
\end{figure}

The results on Figures \ref{fig:subevalgemo} and \ref{fig:subevalemo} evaluate preference. We conclude the subjective evaluation by assessing the perceived expressions elicited by the different lip motion sequences. 
Note that the ability to convey emotional cues in the videos is constrained by the expressiveness of the rendering toolkit Xface (see discussion in Section \ref{sec:discussion}). We only consider the CSG-Emo-Adapted model, which is the expression-aware CSG model with better results according to the objective and subjective comparisons (Table \ref{tab:ll-emo} and Figure \ref{fig:subevalemo}). For this evaluation, we render 10 randomly selected videos per emotional categories (i.e., consensus label). We recruit 32 evaluators each evaluating five pairs of videos, resulting in four evaluations per comparison. For the evaluations, we rely on pairwise comparisons using the same interface shown in Figure \ref{fig:mturk}. The only difference with the previous evaluations is the question, which is rephrased. For example, for happiness, we ask ``Which video looks happier?'' We ask similar questions for anger sadness and frustration. Figure \ref{fig:subemo} gives the results of this evaluation. The results consistently indicate that the lip motion sequences created by the CSG-Emo-Adapted model are selected as more emotional than the CSG model. The preference is statistically significant for happiness (z-test: $p$-value $=0.015$). Using Equation \ref{eq:prop}, we observe that the sequence of the CSG-Emo-Adapted model are selected over the ones from the CSG model 56\% for anger, 65\% for happiness, 57\% for sadness and 48\% for frustration. While the proportion test does not show that these preference are statistical significant (ang: $p$-value $=0.2505$, hap: $p$-value $=0.0676$, sad: $p$-value $=0.7116$), the trend is consistent with the exception of frustration.

\subsection{Discussion}
\label{sec:discussion}

Overall, the experimental evaluations demonstrate that the proposed CSG models perform better than the competitive baselines used in this study. The objective evaluations using log-likelihood of the models reveal the superiority of the expression-aware CSG models over the CSG model, which show the flexibility of the proposed framework to incorporate expressive lip motions. The results using emotion classifiers also show that the emotion expression-aware CSG models are able to generate lip motion sequences conveying emotional cues. 

The results also suggest that the model can be improved. While the  subjective evaluations show a clear trend across emotional classes, the preference toward the expression-aware CSG models are statistically significant only for happiness. An important observation is that some emotions may have a stronger effect on the orofacial area. For example, there is a clear relationship between happiness and the lip configuration. For other emotions, the relationship may be more subtle. We hypothesize that the lack of expressiveness in Xface and the lip parametrization used in this study are the main reasons for not obtaining more decisive results in the subjective evaluation. Xface is a simple toolkit that allows us to render an animation by parametrizing the lip shape using motion capture data, which simplifies our modeling setting. The fact that the emotional cues on the  expression-aware CSG models are not clearly perceived by the evaluators suggest that a more sophisticate rendering toolkit is needed. Furthermore, the IEMOCAP database does not have information for the inner part of the lips, so the lip shape is exclusively defined by the outer lip markers. Therefore, our framework may not capture important lip details that are important to convey the target emotion. We are currently working to address these problems. Even with these limitations, the study clearly demonstrates the modeling potential of the CSG framework, creating exciting opportunities for lip motion generation with speech-driven methods. 

\section{Conclusions}
\label{sec:conclusion}
This paper proposed the CSG model, a conditional GAN that generates orofacial movements from acoustic features. This model learned the conditional distribution of the data with an adversarial training objective, using a generator and a discriminator. The discriminator has to distinguish between real data and samples created by the generator. This adversary training forces the generator to create lip motion trajectories that are realistic. To capture the complex coupling between lip motion and speech, we also presented samples with real audio and motion capture data from different recordings. This type of fake samples presented to the discriminator imposes special emphasis on the temporal dynamic of the lip motion sequences created by the generator. We compared this model with three competitive baselines. The objective and subjective evaluations of the results demonstrated better performance for our model.

One of the strengths of the CSG model is its flexibility to constrain the trajectories by the underlying emotion content, creating expressive lip motion sequences. We proposed two emotion-aware extensions of our model, where we know the target categorical emotion during testing: the CSG-Emo-Adapted, and CSG-Emo-Aware models. The CSG-Emo-Adapted model adapts the network by using the partitions associated with each emotion. The CSG-Emo-Aware model explicitly adds the target emotion as an extra input vector. The results demonstrated that the testing data is better represented by the distribution of the samples generated by the emotion-aware CSG models  than the ones from the CSG model. The emotion classification evaluation using the generated sequences also indicated that both emotion-aware CSG models can generate emotional cues observed in natural recordings. The subjective evaluation showed that the CSG-Emo-Adapted model is perceived more emotional across emotional classes, especially for happiness where the preference was statistically significant.

The experimental evaluation demonstrated the benefits of the proposed CSG models, opening new opportunities to improve the models. The current study focuses on the orofacial area, since this area presents a stronger interplay between articulatory and emotional content. A direct extension of the proposed framework is to generate facial expressions for the entire face, where the emotion can be controlled by specifying the target category. A second extension of the approach is to increase the resolution of the parameters describing the lips. The IEMOCAP corpus does not include inner mouth markers. The inner mouth markers contain subtle differences across emotional categories, which we currently ignore. Using a more dense representation for the lip configuration will help us to generate more expressive and naturalistic animations. Likewise, Xface is not a very expressive toolkit. We expect to create better animations by relying on better rendering toolkits. Finally, the current version of the framework is exclusively driven by speech, without the need for phonetic information. This is one of the key features of our approach. However, if the target application requires better synchronization between lip motion and the phonetic content, the current framework can be extended by constraining the models with the underlying lexical content. For example, we can incorporate lexical content by adding phonemes as additional constraints in the CSG models. We will address these research directions in our future work.




%

\ifCLASSOPTIONcompsoc
\section*{Acknowledgments}
\else
\section*{Acknowledgment}
\fi

This study was funded by the National Science Foundation (NSF) award IIS-1718944.

\ifCLASSOPTIONcaptionsoff
\newpage
\fi




\bibliographystyle{IEEEtranN}
\bibliography{reference}

\end{document}